\documentclass[twocolumn,superscriptaddress,amssymb,amsmath,nobibnotes,aps,prd,showpacs,nofootinbib]{revtex4-1}
\pdfoutput=1

\usepackage{graphicx,bm,color,psfrag,hyperref}
\usepackage{amsfonts}
\usepackage{lipsum}
\usepackage{caption}
\usepackage{subcaption}
\usepackage{newtxtext}
\usepackage{threeparttable,booktabs}

\providecommand{\U}[1]{\protect\rule{.1in}{.1in}}
\newcommand{\be}{\begin{equation}}
\newcommand{\ee}{\end{equation}}

\newcommand{\mincir}{\raise
-3.truept\hbox{\rlap{\hbox{$\sim$}}\raise4.truept\hbox{$<$}\ }}
\newcommand{\magcir}{\raise
-3.truept\hbox{\rlap{\hbox{$\sim$}}\raise4.truept\hbox{$>$}\ }}

\usepackage{color}
\definecolor{darkgreen}{rgb}{0., 0.65, 0.1}

\begin{document}

\title{Constraints on a Bianchi type I spacetime extension of the standard $\Lambda$CDM model}

\author{\"{O}zg\"{u}r Akarsu}
\email{akarsuo@itu.edu.tr}
\affiliation{Department of Physics, \. Istanbul Technical University, Maslak 34469 \. Istanbul, Turkey}

\author{Suresh Kumar}
\email{suresh.kumar@pilani.bits-pilani.ac.in}
\affiliation{Department of Mathematics, BITS Pilani, Pilani Campus, Rajasthan-333031, India}
 
 \author{Shivani Sharma}
\email{p20170025@pilani.bits-pilani.ac.in}
\affiliation{Department of Mathematics, BITS Pilani, Pilani Campus, Rajasthan-333031, India}

\author{Luigi Tedesco}
\email{luigi.tedesco@ba.infn.it}
\affiliation{Dipartimento di Fisica dell'Universit\`a di Bari, 70126 Bari, Italy}
\affiliation{INFN - Sezione di Bari, I-70126 Bari, Italy}

\begin{abstract}
We consider the simplest anisotropic generalization, as a correction, to the standard $\Lambda$CDM model, by replacing the spatially flat Robertson-Walker metric by the Bianchi type-I metric, which brings in a new term $\Omega_{\sigma 0}a^{-6}$ (mimicking the stiff fluid) in the average expansion rate $H(a)$ of the Universe. From Hubble and Pantheon data, relevant to the late Universe ($z\lesssim 2.4$), we obtain the constraint $\Omega_{\sigma0}\lesssim10^{-3}$, in line with the model-independent constraints. When the baryonic acoustic oscillations and cosmic microwave background (CMB) data are included, the constraint improves by 12 orders of magnitude, i.e., $\Omega_{\sigma0}\lesssim10^{-15}$. We find that this constraint could alter neither the matter-radiation equality redshift nor the peak of the matter perturbations. Demanding that the expansion anisotropy has no significant effect on the standard big bang nucleosynthesis (BBN), we find the constraint $\Omega_{\sigma0}\lesssim10^{-23}$. We show explicitly that the constraint from BBN renders the expansion anisotropy irrelevant to make a significant change in the CMB quadrupole temperature, whereas the constraint from the cosmological data in our model provides the temperature change up to $\sim11\, \rm mK$, though it is much  beyond the CMB quadrupole temperature.
\end{abstract}


\maketitle

\section{Introduction}
\label{sec:intro}

The standard Lambda cold dark matter ($\Lambda$CDM) model, relying on the canonical inflationary paradigm \cite{Starobinsky:1980te,Guth:1980zm,Linde:1981mu,Albrecht:1982wi}, is established on the spatially flat Friedmann-Lema\^{i}tre-Robertson-Walker (FLRW) spacetime metric, namely, on the spatially flat  homogeneous, isotropic (viz., maximally symmetric space led by the Copernican principle) and flat Robertson-Walker (RW) spacetime metric for describing the Universe on large scales and general relativity (GR) for describing dynamics of the Universe. The simplest and mathematically tractable step towards a cosmological model with a more realistic background metric is to allow different expansion factors in three orthogonal directions while continuing to demand spatial homogeneity and flatness. It corresponds to replacing the spatially flat RW background metric by the Bianchi type-I background metric  \cite{Collins,Ellis:1998ct}. This, in the absence of any anisotropic source in GR, leads to the generalized Friedmann equation containing \textit{average} Hubble parameter along with a new term, namely, the energy density corresponding to the expansion anisotropy, $\rho_{\sigma}$, scaling as the inverse of the square of the comoving volume. Therefore, $\rho_{\sigma}$ contributes like a stiff or Zeldovich fluid ($p=\rho$) \cite{zel61}, and decreases faster than other known physical sources as the Universe expands \cite{barrow78,Chavanis:2014lra}. Moreover, there is a cosmic no-hair theorem implying that canonical inflation (driven by scalar field model of inflaton) isotropizes the Universe very efficiently \cite{Wald:1983ky,Starobinsky:1982mr}, leaving a residual anisotropy that is negligible for any practical application in the observable Universe. Hence, any high confidence detection of anisotropy in the background expansion of the Universe would have far-reaching consequences on the $\Lambda$CDM model and/or inflationary paradigm and further on the fundamental theories of physics underlying them. Depending on the characteristics of the detected expansion anisotropy, it could be illuminating to the nature of inflaton, dark energy (DE) and even gravitation. For instance, altering the stiff-fluid-like behavior of the expansion anisotropy could be possible mainly by either replacing minimally coupled scalar field models of inflaton or DE by a source yielding anisotropic pressure (e.g., vector fields; see \cite{Barrow:1997sy} for a list of anisotropic stresses and their effects on the expansion anisotropy) or replacing GR by a modified gravity that can give rise to \textit{effective} source yielding anisotropic pressure (e.g., Brans-Dicke theory \cite{Madsen88,Pimentel89,Faraoni:2018qdr}); see, for examples, \cite{Ford:1989me,Barrow:2005qv,Campanelli:2006vb,Koivisto:2007bp,Rodrigues:2007ny,Koivisto:2008xf,Akarsu:2008fj,Watanabe:2009ct,Campanelli2,Campanelli:2009tk,Maleknejad:2012fw,Akarsu:2013dva,Heisenberg:2016wtr,Adak:2016led,Tedesco,Ito:2017bnn,Yang:2018ubt,Akarsu:2019pvi}.

There are various clues for questioning the RW background of the $\Lambda$CDM model. This has been mainly motivated by hints of anomalies in the cosmic microwave background (CMB) distribution first observed on the full sky by the WMAP experiment \cite{deOliveira-Costa:2003utu,Eriksen:2003db,Vielva:2003et,Cruz:2004ce}. These were also observed in the Planck experiment \cite{Ade:2013nlj,Ade:2013ydc,Ade:2013xla,Ade:2013vbw,Schwarz:2015cma}, and followed by many studies, e.g., large angle anomalies \cite{Bennett2}, with the possible clarifications in the alignment of quadrupole and octupole moments \cite{{Land:2005ad},{Ralston:2003pf},{Copi:2003kt}}, the large-scale asymmetry \cite{asymmetry,asymmetry1}, the strange cold spot \cite{Vielva:2003et}, and the low quadrupole moment of the CMB \cite{Bennett11,Campanelli:2006vb,Schwarz:2015cma}. So far, the local deviations from the statistically highly isotropic Gaussianity of the CMB in some directions (the so-called cold spots) could not have been excluded at high confidence levels \cite{Cruz:2004ce,Bennett11,Pontzen10,Ade:2013nlj}. Similarly, it has been shown that the CMB angular power spectrum has a quadrupole power persistently lower than expected from the best-fit $\Lambda$CDM model \cite{Bennett11,Efstathiou:2003tv,Ade:2013kta,Ade:2013zuv,Schwarz:2015cma}. Several explanations for this anomaly have been proposed \cite{DeDeo:2003te,Cline:2003ve,Tsujikawa:2003gh,Koivisto:2005mm,Campanelli:2006vb,Koivisto:2007bp,Rodrigues:2007ny,Gruppuso:2007ya,Campanelli:2007qn,Campanelli2,Campanelli:2009tk,Vazquez:2012}, including the anisotropic expansion of the Universe, that could be developed well after the matter-radiation decoupling, for example, during the domination of DE, say, by means of its anisotropic pressure, acting as a late-time source of not insignificant anisotropy (see, e.g., \cite{Chimento:2005ua,Battye:2006mb,Koivisto:2007bp,Rodrigues:2007ny,Koivisto:2008ig,Koivisto:2008xf,Akarsu:2008fj,Cooray:2008qn,Campanelli2,Campanelli:2009tk,Akarsu:2013dva,Koivisto:2014gia,Koivisto:2015vda,Heisenberg:2016wtr,Yang:2018ubt,Tedesco,Akarsu:2019pvi} for anisotropic DE models and \cite{Mota:2007sz,appleby10,Appleby:2012as,Amendola:2013qna} for constraint studies on the anisotropic DE). On the other hand, in GR in the absence of any anisotropic source as it is the case in the standard $\Lambda$CDM, the CMB provides very tight constraints on the anisotropy at the time of recombination \cite{Martinez95,Bunn:1996ut,Kogut:1997az} of the order of the quadrupole temperature fluctuation $(\Delta T/T)_{\ell=2} \sim 10^{-5}$. And, the stiff-fluid-like behavior of the expansion anisotropy implies an isotropization of the expansion from the recombination up to the present, leading to the typically derived upper bounds on the $\Omega_{\sigma 0}$ today of the order $\sim 10^{-20}$. Thus, any high confidence detection of anisotropic expansion in the present day Universe larger than this expected value within $\Lambda$CDM could be taken as a hint that the late-time Universe is under the influence of some source yielding anisotropic pressure, viz., DE as a source or DE arising as an effective source from a modified gravity.

The implications of the existence of anisotropic expansion in the observable Universe, some of which we pointed out above, led many researchers to study the constraints on the possible anisotropic expansion of the Universe. For example, a direct observational constraint on the expansion anisotropy of the Universe from SN Ia data corresponding to low redshifts is obtained as $\Omega_{\sigma 0}\lesssim 10^{-3}$ \cite{Campanelli:2010zx,Wang:2017ezt}. Because of the stiff-fluid-like redshift dependence of the expansion anisotropy, however, such large upper bounds are not acceptable within the $\Lambda$CDM model, since, in this case, the expansion anisotropy would dominate the Universe just by $z\sim10$ and spoil the standard cosmology for $z\gtrsim10$. There are much stronger constraints such as $\Omega_{\sigma 0}\lesssim10^{-22}$, from Planck CMB temperature and polarization data \cite{Pontzen16,Saadeh:2016bmp,Saadeh:2016sak} and from the light element abundances predicted by big bang nucleosynthesis (BBN) \cite{Barrow:1976rda}. These are model dependent, which assume GR and nonexistence of any anisotropic source so that the stiff-fluid-like behavior of the expansion anisotropy is employed. Recently, the stiff-fluid-like contribution of the expansion anisotropy to the average Hubble parameter is considered in \cite{Hassan1, Hassan2}, and the constraint $\Omega_{\sigma 0}\lesssim10^{-3}$ is obtained using $H(z)$ and/or SN Ia data corresponding to $z\lesssim 2.4$, which is in line with the model-dependent constraints.

In our study, we first present an explicit construction of the Bianchi type-I extension of the $\Lambda$CDM model (Sec. \ref{S2}). Then, to constrain the model and study the Bayesian inference, we consider the most recent Hubble and Pantheon data relevant to the late Universe ($z\lesssim2.4$) and then include the baryonic acoustic oscillations (BAO) and CMB data as well, both of which contain information about the Universe at $z\sim1100$ (see Sec. \ref{S3}). We guarantee expansion anisotropy to remain as a correction all the way to the largest redshifts involved in BAO and CMB data used in our analyses, by fixing the drag redshift as $z_{\rm d}=1059.6$ (involved in BAO analysis) and the last scattering redshift as $z_* = 1089.9$ (involved in CMB data) from the Planck 2015 release for the standard $\Lambda$CDM model \cite{Ade}. We obtain constraints on the Bianchi type-I extension of $\Lambda$CDM in  comparison  with the standard $\Lambda$CDM model by considering different combinations of the datasets, and discuss the results (Sec. \ref{S4}).  We further discuss the results in the context of matter-radiation equality (relevant to the Universe at $z\sim3400$), BBN (relevant to the Universe as $z\sim10^8$), and CMB quadrupole problem (Sec. \ref{S5}). The main findings and conclusions of the study are summarized in Sec. \ref{sec:Conclusions}.

\newpage

\section{Basic Equations and the Model}\label{S2}

The simplest spatially homogeneous but not necessarily isotropic universes can be constructed by the Bianchi type-I spacetime metric \cite{Collins,Ellis:1998ct}, which is the straightforward generalization of the spatially flat FLRW model to allow for different expansion factors in three orthogonal directions, and can be given in matter-comoving (four-velocity being $u^\mu=\delta_0^\mu$) coordinates in the form
\begin{equation}\label{2.1} 
\text{d}s^2=-\text{d}t^2 + A^2(t)\text{d}x^2 + B^2(t)\text{d}y^2 + C^2(t)\text{d}z^2,
\end{equation} 
where $\{A(t),B(t),C(t)\}$ are the directional scale factors along the principal axes $\{x,y,z\}$ and are functions of the cosmic time $t$ only. The corresponding average expansion scale factor is $a(t)=(ABC)^{\frac{1}{3}}$ that arises from the average Hubble parameter defined as $H=\frac{\dot{a}}{a}=\frac{1}{3}\left(H_x + H_y +H_z\right)$. Here, the dot represents the time derivative, and the directional Hubble parameters are defined along the $x$, $y$ and $z$ axes, respectively, as $H_x =\frac{\dot{A}}{A}$, $H_y =\frac{\dot{B}}{B}$ and $H_z =\frac{\dot{C}}{C}$. The most general form of the total energy-momentum tensor $T_{\mu\nu}$ that could be accommodated by this metric is of the form
\begin{equation}\label{2.9}
T^{\nu}_{\,\,\mu} = \text{diag} [-\rho, p_x, p_y, p_z],
\end{equation}
where $\rho$ is the energy density, and $\{p_x, p_y, p_z\}$ are the pressures along the principal axes $\{x,y,z\}$. In view of Eqs. \eqref{2.1} and \eqref{2.9}, Einstein's field equations
\begin{equation}\label{2.8}
R_{\mu\nu}-\frac{1}{2}g_{\mu\nu} R = 8\pi GT_{\mu\nu},
\end{equation}
where $R_{\mu\nu}$ is the Ricci tensor, $R$ is the Ricci scalar, $g_{\mu\nu}$ is the metric tensor and $G$ is Newton's gravitational constant, yield the following set of differential equations:
\begin{align}
H_x H_y + H_y H_z + H_z H_x &= 8\pi G \rho,\label{2.13}\\
-\dot{H_y} - H_y^2 - \dot{H_z} - H_z^2 - H_y H_z &=8\pi G p_x ,\label{2.11}\\
-\dot{H_z} - H_z^2 - \dot{H_x} - H_x^2 - H_z H_x &=8\pi G p_y,\label{2.12}\\
-\dot{H_x} - H_x^2 - \dot{H_y} - H_y^2 - H_x H_y &=8\pi G p_z.\label{2.10}
\end{align}
This set of equations satisfies $T^{\mu\nu}_{\;\;\;;\nu}=0$ (the conservation equation for the total energy-momentum tensor representing all sources in the Universe) via $G^{\mu\nu}_{\;\;\;;\nu}=0$ as a consequence of the second Bianchi identity. This leads to the continuity equation
\begin{equation}\label{2.18}
\dot{\rho}+3H\rho+H_x p_x+H_y p_y+H_z p_z =0.
\end{equation}

We intend to investigate the simplest anisotropic generalization, as a correction, to the base $\Lambda$CDM model. We replace the spatially flat RW background metric of the standard $\Lambda$CDM cosmology by the Bianchi type-I background metric while keeping the physical ingredients of the Universe as usual, summarized as follows: We consider the pressureless fluid or dust (CDM, baryons) described by the equation of state (EOS): $p_{\rm m}/\rho_{\rm m}=0$, radiation (photons $\gamma$, neutrinos $\nu$) described by the EOS: $p_{\rm r}=\rho_{\rm r}/3$, and DE mimicked by the cosmological constant $\Lambda$ described by the EOS: $p_{\Lambda}=-\rho_{\Lambda}$. These all yield isotropic pressure (i.e., $p_x=p_y=p_z=p$), reducing the continuity equation \eqref{2.18} to $\dot{\rho} + 3H(\rho+p) =0$. We assume that these sources interact only gravitationally so that the continuity equation is satisfied separately by each source, and this leads to
\begin{equation}\label{2.19}
\rho_\text{r} = \rho_{\text{r}0}a^{-4},\,\,
\rho_\text{m} = \rho_{\text{m}0} a^{-3}\,\,\, \text{and}\,\,\,
\rho_\Lambda = \text{const},
\end{equation}
for which $a=1$ corresponds to the present time of the Universe. Here and onward, a subscript 0 attached to any quantity implies its value in the present time Universe.  We consider the radiation content, $\rho_{\rm r}=\rho_{\gamma}+\rho_{\nu}$, by including three neutrino species ($N_{\rm eff}=3.046$) with minimum allowed mass $\sum m_{\nu}=0.06\, {\rm eV}$, theoretically well determined within the framework of the standard model of particle physics. The photon energy density today $\rho_{{\gamma}0}$ is well constrained. For, it has a simple relation $\rho_{\gamma}=\frac{\pi^2}{15} T_{\rm CMB}^4$  with the CMB monopole temperature (see \cite{Dodelson03} for further details), which today is very precisely measured: $T_{{\rm CMB}0}=2.7255\pm 0.0006\,{\rm K}$ \cite{Fixsen09}. The density parameter of radiation is $\Omega_{\rm r0}=2.469\times 10^{-5} h^{-2}(1+0.2271 N_{\rm eff})$, where $h=H_0/100\, {\rm km\,s}^{-1}\,{\rm Mpc}^{-1}$ is the dimensionless Hubble constant \cite{Komatsu:2010fb}.


Next, we need to find the contribution of the expansion anisotropy to the anisotropic $\Lambda$CDM model, which could be quantified through the shear scalar
\begin{equation}
\sigma^2 \equiv \frac{1}{2}\sigma_{\alpha \beta} \, \sigma^{\alpha \beta},
\end{equation}
where $\sigma_{\alpha \beta} = \frac {1} {2} (u_{\mu;\nu}+u_{\nu; \mu})h^{\mu}_{\;\alpha}h^{\nu}_{\;\beta} - \frac{1}{3} u^{\mu}_{\; ;\mu} \, h_{\alpha \beta}$ is the shear tensor. Here $h_{\mu \nu} = g_{\mu \nu} + u_{\mu} u_{\nu}$ is the so-called ``projection tensor" with $u_{\mu}$ being the four-velocity in the comoving coordinates \cite{Ellis:1998ct}. For the Bianchi type-I spacetime metric \eqref{2.1}, $\sigma^2$ is obtained in terms of the directional Hubble parameters as
\begin{equation}\label{sigma2}
\sigma^2 = \frac{1}{6}\left[(H_x -H_y)^2 + (H_y - H_z)^2 + (H_z -H_x)^2\right].
\end{equation}
From \eqref{2.11} -- \eqref{2.10}, in the absence of any anisotropic source in line with aforementioned standard cosmological sources, one can obtain
\begin{equation}
\frac{H_x -H_y}{x_1}=\frac{H_y -H_z}{x_2}=\frac{H_z -H_x}{x_3}=a^{-3},
\end{equation}
where $x_1$, $x_2$ and $x_3$ are integration constants. Then \eqref{sigma2} reduces to
\begin{equation}\label{2.17}
\sigma^2 = \sigma_0^2 a^{-6},
\end{equation}
where $\sigma_0^2=\frac{1}{6}(x_1^2 + x_2^2 + x_3^2)$. The density parameter corresponding to the shear scalar can be defined as follows:
\begin{equation}\label{2.7a}
\Omega_{\sigma}=\frac{\sigma^2}{3H^2},
\end{equation}
which quantifies the expansion anisotropy through its contribution to the average expansion rate $H(z)$ of the Universe in line with the other density parameters.

Finally, the Friedmann equation \eqref{2.13} for the anisotropic $\Lambda$CDM model can be recast as follows:
\begin{equation}\label{2.20}
3H^2=8\pi G(\rho_{\rm r}+\rho_{\rm m}+\rho_{\Lambda})+\sigma^2.
\end{equation} 
Using \eqref{2.19}, \eqref{2.17} and \eqref{2.7a}, it leads to
\begin{equation}\label{model}
\frac{H^2(a)}{H_0^2} =\Omega_{\sigma0} a^{-6} +\Omega_{\text{r}0}a^{-4} + \Omega_{\text{m}0}a^{-3} + \Omega_{\Lambda0},
\end{equation}
where $\Omega_{i0}=\rho_{i0}/\rho_{c0}$ is the present day density parameter of the $i$th fluid, $\rho_{c0}=\frac{3H_0^2}{8\pi G}$ being the present critical density of the Universe, and $\Omega_{\sigma0}=\frac{\sigma_0^2}{3H_0^2}$ from \eqref{2.7a}. We note that $\Omega_{\sigma0} + \Omega_{\text{r}0}+ \Omega_{\text{m}0}+\Omega_{\Lambda0}=1$. Further, this Friedmann equation \eqref{model} obtained within Bianchi type-I spacetime differs from that of the base $\Lambda$CDM with its two aspects: (i) Here $H(z)$ is the average expansion rate, and that the expansion rates along the different principal axes---$H_x$, $H_y$ and $H_z$---need not necessarily be the same. Accordingly, we define the corresponding average redshift $z$ through the average scale factor $a$ as $z=-1+\frac{1}{a}$. (ii) There is a new term $\Omega_{\sigma 0}a^{-6}$ on the top of the base $\Lambda$CDM model, which quantifies the contribution of the expansion anisotropy to the average expansion rate. The shear scalar $\sigma^2$ [see \eqref{2.17}] contributes to the Friedmann equation like a stiff fluid $\rho_{\rm s}\propto a^{-6}$ described by the EOS of the form $p_{\rm s}/\rho_{\rm s}=1$, when stiff or Zeldovich fluid \cite{zel61} is included to the base $\Lambda$CDM model. \footnote{A detailed theoretical investigation of the standard $\Lambda$CDM model augmented by stiff fluid was recently presented in \cite{Chavanis:2014lra}. One may check that these two models have mathematically exactly the same Friedmann equation, though they are physically different. However, at the background level, the observational constraints obtained in this study are valid for that model too.} This is a generic result for general relativistic Bianchi type-I cosmologies in the absence of any kind of anisotropic source. \footnote{The presence of anisotropic sources or modifications to GR leading to effective sources yielding anisotropic pressure alters the $\sigma^2 \propto a^{-6}$ relation; see Sec. \ref{sec:Conclusions} for further comments.}

\section{Bayesian Inference}\label{S3}

In recent years, Bayesian inference has been extensively used
in parameter estimation and model comparison in cosmological studies \cite{bethoven, simony, maria, uendert, antonella}. According to Bayes' theorem, the posterior distribution
$P(\Theta|D, M)$  of the parameters $\Theta$ of a given model $M$ is written as
\begin{equation}\label{bayes}
		P(\Theta|D, M) = \frac{\mathcal{L}(D|\Theta, M) \pi(\Theta|M)}{\mathcal{E}(D|M)},
\end{equation}
where $D$ is the cosmological data, $\mathcal{L}(D|\Theta, M)$ is the likelihood, $\pi(\Theta|M)$ is the prior probability of the model parameters, and $\mathcal{E}(D|M)$ is the Bayesian evidence, given by
\begin{equation}\label{evidence}
\mathcal{E}(D|M) = \int_M \mathcal{L}(D|\Theta, M) \pi(\Theta|M) d\Theta.
\end{equation}
For parameter estimation in cosmological models, it is common to use a multivariate Gaussian likelihood given by
\begin{equation}
	\mathcal{L}(D|\Theta,M) \propto \exp \Bigg[-\frac{\chi^{2}(D|\Theta,M)}{2}\Bigg],
\end{equation}
where $\chi^{2}(D|\Theta,M)$ is the chi-squared function for the dataset $D$. In case of uniform prior distribution $\pi(\Theta|M)$ of the model parameters, Eq. \eqref{bayes} leads to
\begin{equation}
P(\Theta|D, M)\propto \exp \Bigg[-\frac{\chi^{2}(D|\Theta,M)}{2}\Bigg].
\end{equation}
Thus, the posterior probability $P(\Theta|D, M)$ or the likelihood $\mathcal{L}(D|\Theta,M)$ is maximum where the $\chi^{2}(D|\Theta,M)$ is minimum.

In order to compare a model $M_i$ with a reference model $M_j$, the ratio of the posterior probabilities of the two models, is computed by using \cite{trotta}
\begin{equation}
\frac{P(M_i|D)}{P(M_j|D)} = B_{ij}\frac{P(M_i)}{P(M_j)},
\end{equation}
where $B_{ij}$ is the Bayes' factor, defined as
\begin{equation}\label{bayes_factor}
B_{ij} = \frac{\mathcal{E}_i}{\mathcal{E}_j}.
\end{equation}

The Bayes' factor is commonly interpreted using the Jeffrey's scale \cite{Jeffreys}, given in Table \ref{tab:BE}. This table suggests that the evidence in favor of or against the model $M_{i}$ relative to model $M_{j}$ is weak or inconclusive in case $ |\ln {B_{ij}}|< 1$. Further, the reference model $M_{j}$  is favored over the model $M_i$ when $\ln{B_{ij}} < -1$. In our analysis, we will adopt the $\Lambda$CDM model as the reference model $M_{j}$.

\begin{table} [htb!]
\begin{center}
	\caption {Jeffrey's scale} \label{tab:BE}
    \begin{tabular}{c | c}
        \hline
        $|\ln B_{ij}|$ & Strength of evidence\\
        \hline
        $[0,1)$ &
          Weak/inconclusive\\
           $[1,3)$ &
          Positive/definite\\
        $[3,5)$ &
          Strong\\
        $\geq 5$ &
           Very strong\\
          \hline 
        
      \end{tabular}
 \end{center}
\end{table}


\subsection {Data and likelihoods}

\subsubsection{$H(z)$}
We consider a compilation of 36 $H(z)$ measurements as shown in Table \ref{tab:H_z}, viz., the first 31 measurements obtained from the cosmic chronometric method \cite{Morescoetal2012}, three correlated measurements (at $z=0.38$, $z=0.51$ and $z=0.61$) from the BAO signal in galaxy distribution \cite{Alametal2017}, and the last two measurements (at $z=2.34$ and $z=2.36$) determined from the BAO signal in Ly-$\alpha$ forest distribution alone or cross-correlated with quasistellar objects (QSOs) \cite{Delubacetal2015, Font-Riberaetal2014}. 

The chi-squared function for the 33 $H(z)$ measurements, denoted by $\chi^2_{\rm CC+Ly{\alpha}}$, is
\begin{equation}
\chi^2_{\rm CC+Ly{\alpha}} = \sum_{i=1}^{33} \frac{[H^{\text{obs}}(z_i)-H^{\text{th}}(z_i)]^2}{\sigma^2_{H^{\text{obs}}(z_i)}},
\end{equation}
where $H^{\text{obs}}(z_i)$ is the observed value of the Hubble parameter with the standard deviation $\sigma^2_{H^{\text{obs}}(z_i)}$ as given in the Table \ref{tab:H_z}, and $H^\text{th}(z_i)$ is the theoretical value obtained from the cosmological model under consideration.

On the other hand, the covariance matrix related to the three measurements from galaxy distribution \cite{Alametal2017} reads
\begin{equation}\label{23a}
 C=
  \left[ {\begin{array}{ccc}
   3.65 & 1.78 & 0.93 \\
   1.78 & 3.65 & 2.20 \\
   0.93 & 2.20 & 4.45 \\
  \end{array} } \right].
\end{equation}
The chi-squared function for the three galaxy distribution measurements is 
\begin{equation}
\chi^2_{\rm Galaxy} = M^T C^{-1} M,
\end{equation}
where 
\begin{equation}\label{23b}
 M=
  \left[ {\begin{array}{c}
   H^\text{obs}(0.38)-H^\text{th}(0.38) \\
   H^\text{obs}(0.51)-H^\text{th}(0.51) \\
   H^\text{obs}(0.61)-H^\text{th}(0.61) \\
  \end{array} } \right].
\end{equation}

\begin{table}[!ht]
\begin{center}
\footnotesize
\caption{Hubble parameter data.}\label{tab:H_z}
\begin{tabular}{l|c|c|c}
  \hline
  \hline
$z_i$ &   $H^{\text{obs}}(z_i)$ [km s$^{-1}$ Mpc$^{-1}$]    &  $\sigma_{H^{\text{obs}}(z_i)}$ & Reference \\ \hline
0.07	&	69	&	19.6	& 	\cite{Zhangetal2014}\\
0.09	&	69	&	12	& 	\cite{Simonetal2005} \\
0.12	&	68.6 &	26.2	& 	\cite{Zhangetal2014} \\
0.17	&	83	&	8	& 	\cite{Simonetal2005}\\
0.179	&	75	&	4	& 	\cite{Morescoetal2012} \\
0.199	&	75	&	5	& 	\cite{Morescoetal2012}\\
0.2	    &	72.9 &	29.6	&  \cite{Zhangetal2014} \\
0.27	&	77	&	14	&   \cite{Simonetal2005} \\
0.28	&	88.8 &	36.6	&   \cite{Zhangetal2014} \\
0.352	&	83	&	14	&  \cite{Morescoetal2012}\\
0.38	&	81.9 &	1.9	&   \cite{Alametal2017}\\
0.3802	&	83	&	13.5	&  \cite{Morescoetal2016} \\
0.4	        &	95	&	17	&   \cite{Simonetal2005} \\
0.4004	&	77	&	10.2	&   \cite{Morescoetal2016} \\
0.4247	&	87.1	&	11.2	&  \cite{Morescoetal2016} \\
0.4497	&	92.8	&	12.9	&   \cite{Morescoetal2016}\\
0.47        &       89  &   50      &  \cite{Ratsimbazafyetal2017} \\
0.4783	&	80.9	&	9	&   \cite{Morescoetal2016}\\
0.48	&	97	&	62	&  \cite{Ratsimbazafyetal2017} \\
0.51	&	90.8	&	1.9	&   \cite{Alametal2017} \\
0.593	&	104	&	13	&   \cite{Morescoetal2012} \\
0.61	&	97.8	&	2.1	&   \cite{Alametal2017} \\
0.68	&	92	&	8	&   \cite{Morescoetal2012} \\
0.781	&	105	&	12	&  \cite{Morescoetal2012} \\
0.875	&	125	&	17	&   \cite{Morescoetal2012} \\
0.88	&	90	&	40	&    \cite{Ratsimbazafyetal2017} \\
0.9	        &	117	&	23	&   \cite{Simonetal2005} \\
1.037	&	154	&	20	&  \cite{Morescoetal2012} \\
1.3	        &	168	&	17	&    \cite{Simonetal2005} \\
1.363	&	160	&	33.6	&  \cite{Moresco2015}\\
1.43	&	177	&	18	&   \cite{Simonetal2005} \\
1.53	&	140	&	14	&  \cite{Simonetal2005} \\
1.75	&	202	&	40	&   \cite{Simonetal2005} \\
1.965	&	186.5	&	50.4 &   \cite{Moresco2015} \\
2.34	&	223 &	7	&  \cite{Delubacetal2015} \\
2.36	&	227	&	8	&   \cite{Font-Riberaetal2014} \\ \hline
\end{tabular}
\end{center}
\end{table}
Henceforth, the combined chi-squared function for $H(z)$ measurements, denoted by $\chi^2_{\rm H}$, is
\begin{equation}
    \chi^2_{\rm H} = \chi^2_{\rm CC+Ly{\alpha}} + \chi^2_{\rm Galaxy}.
\end{equation}


\subsubsection{BAO}
BAO measurements are useful to study the angular-diameter distance as a redshift function and the evolution of the Hubble parameter. 
These measurements are represented by using angular scale and redshift separation. They are commonly written in terms of the dimensionless ratio 
\begin{equation}
d(z)=\frac{r_{\rm s}(z_{\rm d})}{D_{\rm V}(z)},
\end{equation}
where $r_{\rm s}(z_{\rm d})$ represents the comoving size of the sound horizon at the drag redshift, $z_{\rm d}=1059.6$ \cite{Ade}:
\begin{equation}
\label{comoving size}
r_{\rm s}(z_{\rm d})=\int_{z_{\rm d}}^\infty \frac{c_{\rm s}dz}{H(z)}. 
\end{equation}
Here, $c_{\rm s}=\frac{c}{\sqrt{3(1+\mathcal{R})}}$ represents the sound speed of the baryon-photon fluid, and $\mathcal{R}=\frac{3\Omega_{\rm b0}}{4\Omega_{\rm r0}(1+z)}$ with $\Omega_{\rm b0}=0.022h^{-2}$ \cite{Cooke} and $\Omega_{\rm r0} =\Omega_{\gamma 0}\Big(1+\frac{7}{8}(\frac{4}{11})^{\frac{4}{3}}N_{\rm eff}\Big)$, where $\Omega_{\gamma 0}=2.469\times 10^{-5}h^{-2}$ and $N_{\rm eff}= 3.046$ \cite{Dodelson03}.

Further, $D_{\rm V}(z)$ is the volume averaged distance that gives the relation between the line of sight and transverse distance scale \cite{BAO, BAO1}:
\begin{equation}
D_{\rm V}(z)=\Bigg[(1+z)^2D_{\rm A}(z)^2\frac{cz}{H(z)}\Bigg]^{1/3},
\end{equation}
where $D_{\rm A}(z)=\frac{c}{1+z}\int_0^z\frac{dz}{H(z)}$ is the angular diameter distance, and $c$ is the speed of light.

\begin{table}
	\begin{center}
	\caption{\label{tab:BAO}BAO data}
		\begin{tabular}{l c c c c}
		\hline
			Survey     & \multicolumn{1}{c}{$z_i$} & \multicolumn{1}{c}{$d(z_i)$} & \multicolumn{1}{c}{$\sigma_{d(z_i)}$} & Ref. \\
			\hline
			6dFGS      & $0.106$ & $0.3360$ & $0.0150$ & \cite{bao1s} \\
			MGS        & $0.15$  & $0.2239$ & $0.0084$ & \cite{bao2} \\
			BOSS LOWZ  & $0.32$  & $0.1181$ & $0.0024$ & \cite{bao4} \\
			SDSS(R)    & $0.35$  & $0.1126$ & $0.0022$ & \cite{bao3} \\
			BOSS CMASS & $0.57$  & $0.0726$ & $0.0007$ & \cite{bao4} \\
			WiggleZ    & $0.44$  & $0.073$  & $0.0012$ & \cite{Wigglez}\\
			WiggleZ    & $0.6$   & $0.0726$ & $0.0004$ & \cite{Wigglez}\\
			WiggleZ    & $0.73$  & $0.0592$ & $0.0004$ & \cite{Wigglez}\\
			\hline
		\end{tabular}
		\end{center}
\end{table}

The chi-squared function of BAO measurements from the first five surveys as mentioned in Table \ref{tab:BAO}, denoted by $\chi^2_{\rm NW}$, reads 
\begin{equation}
\chi^2_{\rm NW}=\sum_{i=1}^{5}\Bigg[\frac{d^{\text{obs}}(z_i)-d^{\text{th}}(z_i)}{\sigma_{d(z_i)}}\Bigg]^2,
\end{equation}
where $d^{\text{obs}}(z_i)$ is the observed value of the dimensionless ratio with the uncertainty $\sigma_{d(z_i)}$ as given in the Table \ref{tab:BAO} and  $d^{\text{th}}(z_i)$ is the theoretical value obtained from the cosmological model under consideration.

We shall also consider the three data points from the WiggleZ survey \cite{Wigglez}. The inverse covariance matrix related with these data points is given by 
\begin{equation}\label{33}
 C^{-1}=
  \left[ {\begin{array}{ccc}
   1040.3 & -807.5 & 336.8 \\
   -807.5 & 3720.3 & -1551.9 \\
   336.8 & -1551.9 & 2914.9 \\
  \end{array} } \right].
\end{equation}

For the WiggleZ data, the chi-squared function, denoted by $\chi^2_{\rm W}$, is defined as
\begin{equation}
\chi^2_{\rm W} = D^T C^{-1} D,
\end{equation}
where
\begin{equation}\label{34}
 D=
  \left[ {\begin{array}{c}
   d^\text{obs}(0.44)-d^\text{th}(0.44) \\
   d^\text{obs}(0.6)-d^\text{th}(0.6) \\
   d^\text{obs}(0.73)-d^\text{th}(0.73) \\
  \end{array} } \right].
\end{equation}

Thus, the chi-squared function for the total BAO contribution, denoted by $\chi^2_{\rm BAO}$, gives
\begin{equation}
\chi^2_{\rm BAO}=\chi^2_{\text{NW}} +\chi^2_{\rm W}.
\end{equation}


\subsubsection{CMB}
From the compressed likelihood information of Planck 2015 CMB data \cite{Ade}, we use the angular scale of the sound horizon at the last scattering surface, denoted by $l_{\rm a}$, defined as
\begin{equation}
    l_{\rm a} =\pi\frac{r(z_*)}{r_{\rm s}(z_*)},
\end{equation}
where $r(z_*)$ is the comoving distance to the last scattering surface, evaluated as
\begin{equation}\label{comoving distance}
r(z_*)=\int_0^{z_*}\frac{c\text{d}z}{H(z)},
\end{equation}
and $r_{\rm s}(z_*)$ is the size of the comoving sound horizon [see \eqref{comoving size}] evaluated at $z_* = 1089.9$, the redshift of last scattering \cite{Ade}.

 
The chi-squared function of CMB, denoted by $\chi^2_{\rm CMB}$, reads
 \begin{equation}
\chi^2_{\rm CMB} =\frac{\big(l_{\rm a}^{\text{obs}}-l_{\rm a}^{\text{th}}\big)^2}{\sigma^2_{l_{\rm a}}},
 \end{equation}
where $l_{\rm a}^{\text{obs}}= 301.63 $ is the observed value of the angular scale of the sound horizon with uncertainty $\sigma_{l_{\rm a}}= 0.15$ (see \cite{Ade}) and $l_{\rm a}^{\text{th}}$ is the theoretical value obtained from the cosmological model under consideration.

\subsubsection{Pantheon supernovae type Ia}
The Pantheon sample is a combination of five subsamples: PS1, SDSS, SNLS, low-$z$, and HST that gives the largest supernovae sample of 1048 measurements, spanning over the redshift range: $0.01<z<2.3$ \cite{P1}. 
Following \cite{P1}, we use Pantheon data in line with the joint light-curve sample \cite{P2} but ignoring the stretch luminosity parameter $\alpha$ and the color luminosity parameter $\beta$. 

%
 
The theoretical distance modulus is defined by \cite{P2} 
\begin{equation}
\mu_\text{th}=   5\log_{10} \frac{d_L(z_{\text{hel}},z_{\text{cmb}})}{\text{10\,pc}},
\end{equation}
where $d_L$ is the luminosity distance 
given by $d_L=(c/H_0)D_L$. Further,
\begin{equation}
D_{\rm L}= (1+z_{\rm hel})\int_0^{z_{\rm cmb}}\frac{H_0 dz}{H(z)},
\end{equation}
where $z_{\rm hel}$ is the heliocentric redshift and $z_{\rm cmb}$ is the redshift of the CMB rest frame.

The observed distance modulus \cite{P2} is given by
\begin{equation}
\mu_\text{obs}=m_{\rm B} -\mathcal{M},
\end{equation}
where $m_{\rm B}$ is the observed peak magnitude in the rest frame of the $\rm B$ band and $\mathcal{M}$ is the nuisance parameter.

In the case of Pantheon data, the chi-squared function, denoted by $\chi^2_{\rm Pan}$, is given by
\begin{equation}
\chi^2_{\text{Pan}}=m^TC^{-1}m,
\end{equation}
where $C$ is the covariance matrix of $\mu_{\text{obs}}$ given in \cite{P3} and $m=m_B-m_{\text{th}}$ with 
\begin{equation}
m_{\text{th}}=5\log_{10}D_L +\mathcal{M}.
\end{equation}

The total covariance matrix as in \cite{P1} reads
\begin{equation}
C=D_{\text{stat}} + C_{\text{sys}},
\end{equation}
where ${C}_{\text{sys}}$ is the systematic covariance matrix and ${D}_{\text{stat}}$ is the diagonal covariance matrix of the statistical uncertainty, calculated as
\begin{equation}
{D}_{\text{stat}, ii} = \sigma^{2}_{m_{\rm B, i}}.
\end{equation}
The systematic covariance matrix together with $m_{\rm B, i}$, $\sigma^{2}_{m_{\rm B, i}}$, $z_{\text{cmb}}$, $z_{\text{hel}}$ for the $i{\text{th}}$ SN Ia are available in \cite{P1}.

\subsection{Methodology}
To obtain observational constraints on the anisotropic $\Lambda$CDM model parameters from the above-mentioned $H(z)$, BAO, CMB and Pantheon data, we use PyMultiNest \cite{Buchner} code, which is a Python interface for MultiNest
\cite{feroz1,feroz2,feroz3}, and a generic Bayesian inference tool that uses the nested sampling \cite{Skilling} to calculate the Bayesian evidence,
and also allows for parameter inference.

We consider multivariate joint Gaussian likelihood given by 
\begin{equation}\label{27}
    \mathcal{L}_{\rm Joint} \propto \text{exp}\left(\frac{-\chi^2_{\rm Joint}}{2}\right),
\end{equation}
where the joint chi-squared function of all the datasets reads
\begin{equation}
\chi^2_{\rm Joint} = \chi^2_{\rm H}+\chi^2_{\rm BAO} +\chi^2_{\rm CMB} + \chi^2_{\rm Pan}.
\end{equation}

In our study, we choose uniform prior distribution for all the model parameters $H_0$, $\Omega_{\rm m0}$ and $\Omega_{\sigma0}$, viz., $55\leq H_0\leq 85$, $0.1\leq \Omega_{\rm m0}\leq 0.5$ and $0\leq \Omega_{\sigma0}\leq0.1$, respectively. In the case of the data combinations with CMB and/or BAO, we have chosen the prior range $0\leq \Omega_{\sigma0}\leq 10^{-14}$. \footnote{CMB and BAO data likelihoods, in our study, use the fixed high redshifts such as the drag redshift $z_d$ and the last-scattering redshift $z_*$, and therefore we expect a small amount of anisotropy as a correction on the top of standard $\Lambda$CDM evolution of the Universe in our model and results (for more details see Sections I and IV). The Universe should be matter dominated at the recombination that takes place at $z\sim1100$, which is physically closely related to the last scattering surface redshift $z_*$ and drag redshift $z_{\rm d}$. Accordingly, using $\Omega_{\rm m}(z\sim1100)\approx 1$ and, say, $\Omega_{\sigma}(z\sim1100)\lesssim 10^{-2}$ into $\frac{\Omega_{\sigma}}{\Omega_{\rm m}}=\frac{\Omega_{\sigma 0}}{\Omega_{\rm m0}}(1+z)^{3}$, we find that the upper bound for $\frac{\Omega_{\sigma 0}}{\Omega_{\rm m0}}$ should be $\sim 10^{-11}$. Starting from this upper bound, during test runs of the code, we found that the prior range $0\leq \Omega_{\sigma0}\leq 10^{-14}$ is good enough to extract the information about $\Omega_{\sigma 0}$.  }


\section{Results and Discussion}\label{S4}

\begin{figure}[hbt!]
	\centering
	\includegraphics[width=8cm]{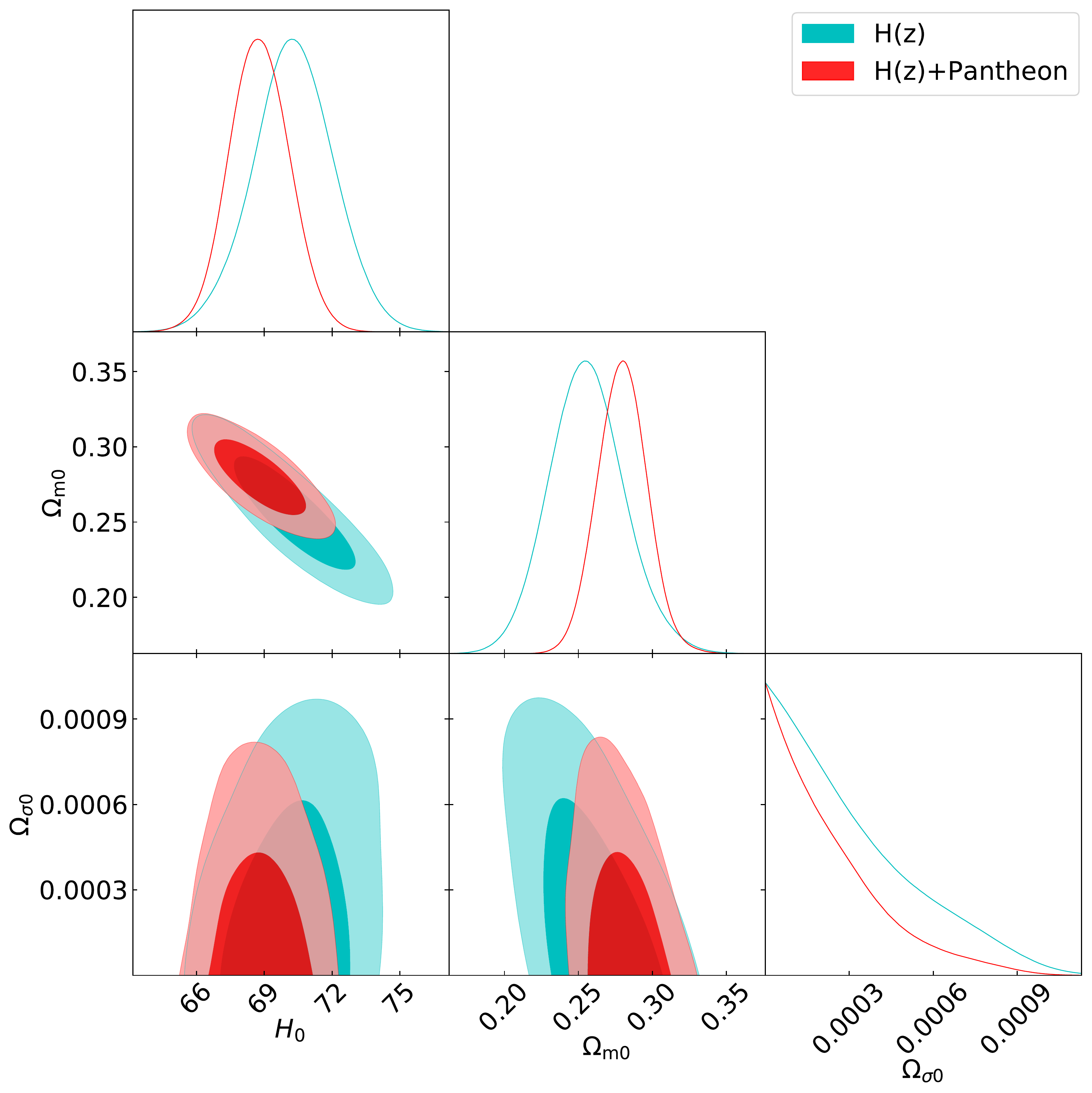}
	\caption{One-dimensional and two-dimensional marginalized confidence regions (68\% and 95\%  C.L.) for the anisotropic $ \Lambda$CDM parameters $H_0$, $\Omega_{\rm m0}$ and $\Omega_{\sigma 0}$ from $H(z)$ and $H(z)$ + Pantheon data.}
	\label{fig:comblowredshift}
\end{figure}

Table \ref{Results} displays the constraints on the parameters of the anisotropic $\Lambda$CDM model in comparison to the base $\Lambda$CDM model from two relatively low redshift datasets: $H(z)$ and $H(z)$ + Pantheon. Figure \ref{fig:comblowredshift} shows the one-dimensional and two-dimensional (68\% and 95\%) confidence regions of the anisotropic $\Lambda$CDM model parameters for the two datasets. In both cases, we see that the upper bound on the anisotropy parameter $\Omega_{\sigma 0}$ is of the order $10^{-3}$ in line with the model-independent constraints, e.g., from type Ia supernovae data \cite{Campanelli:2010zx,Wang:2017ezt}. Also, this order is similar to the one obtained in \cite{Hassan1, Hassan2}. However, such an amount of expansion anisotropy in the present Universe within the anisotropic $\Lambda$CDM model implies the domination of the anisotropy in the Universe by $z\sim 10$. This would lead to large deviations from the standard $\Lambda$CDM model and spoil the successful description of the Universe for $z\gtrsim 10$. Hence, we see that the upper bound on the density parameter corresponding to the present day expansion anisotropy at the level $10^{-3}$ may not be realistic within the anisotropic $\Lambda$CDM model. 

In this study, we aim to constrain the allowed amount of anisotropy from the observational data on the top of the standard $\Lambda$CDM model, so that on average the standard $\Lambda$CDM Universe is not spoiled. Notice that the new term, i.e., the expansion anisotropy, is the fastest growing term with the increasing $z$ in $H(z)$. Therefore, for guaranteeing expansion anisotropy to remain as a correction all the way to the largest redshifts involved in BAO and CMB data used in our analyses here, we fix the drag redshift as $z_{\rm d}=1059.6$ (involved in BAO analysis) and the last scattering redshift as $z_* = 1089.9$ (involved in CMB data), where the fixed values are taken from the Planck 2015 release for the standard $\Lambda$CDM model \cite{Ade}. With these settings, when we include CMB and/or BAO data, $\Omega_{\sigma 0}$ is of the order $10^{-15}$ in all cases (see Table \ref{comb_results}). This shows that the CMB and/or BAO data offer tight constraints on $\Omega_{\sigma 0}$. The reason is that $H(z)$ and Pantheon data correspond to low redshifts and are therefore unable to put tight constraints on $\Omega_{\sigma 0}$. On the other hand, CMB and BAO data likelihoods include fixed high redshifts such as the drag redshift $z_d$ and the last scattering redshift $z_*$ and, therefore, preserve the standard $\Lambda$CDM evolution of the Universe at early times. In other words, $10^{-15}$ is the allowed order of anisotropy on the top of standard $\Lambda$CDM evolution of the Universe in our results. In what follows, we discuss the results with the constraints obtained in Table \ref{comb_results}.

\begin{table}[t]
\caption{\label{Results} Constraints (68\% and 95\% C.L.) on the anisotropic $\Lambda$CDM and $\Lambda$CDM model parameters from the $H(z)$ and Pantheon data.}
   \centering
    {%
    \begin{tabular}{c|c|c}
        \hline
        Data  & $H(z)$  & $H(z)$+Pantheon   \\ 
        \hline
        \hline
         Anisotropic $\Lambda$CDM & \\
        \hline
        $H_0$ &
        $70.4^{+1.9+3.5}_{-1.9-3.8}$ &
        $68.7^{+1.3+2.5}_{-1.3-2.7}$ \\
        
        $\Omega_{\rm m0}$ & 
        $0.254^{+0.024+0.054}_{-0.028-0.051}$ &
        $0.281^{+0.017+0.034}_{-0.017-0.034}$\\
        
        $\Omega_{\sigma 0}$ (95$\%$ C.L.) &
        $<4.6 \times 10^{-4}$ &
        $<7.4 \times 10^{-4}$\\

        $\ln \mathcal{E}$ & 
        $-15.29 \pm 0.09$ & 
        $-536.11 \pm 0.16$ \\
        
        \hline
        \hline

        $\Lambda$CDM &\\
        \hline
        $H_0$ &
        $69.6^{+1.8+3.5}_{-1.8-3.4}$ &
        $68.8^{+1.4+2.7}_{-1.4-2.6}$ 
        \\
        
        $\Omega_{\rm m0}$ &
        $0.271^{+0.023+0.048}_{-0.025-0.046}$ &
        $0.285^{+0.017+0.035}_{-0.017-0.032}$ 
        \\
        
        $\ln \mathcal{E}$ & 
        $-14.64 \pm 0.09$ & 
        $-535.24 \pm 0.16$ \\
        
        \hline
        \hline
        
    \end{tabular}%
    }
    
\end{table}

\begin{table*}[htb] 

\caption{\label{comb_results} Constraints (68\% and 95\% C.L.) on the anisotropic $\Lambda$CDM and $\Lambda$CDM model parameters from four different combinations of $H(z)$, CMB, BAO and Pantheon data. The Bayesian evidence is also displayed in each case.}
    \centering
    {%
    \begin{tabular}{c|c|c|c|c}
        \hline
        Parameter & $H(z)$+CMB & $H(z)$+BAO+CMB & $H(z)$+Pantheon+CMB  & $H(z)$+BAO+Pantheon+CMB  \\ 
        \hline
        \hline
          Anisotropic $\Lambda$CDM & & & \\
        \hline
        $H_0$ &
        $70.4^{+1.7+3.3}_{-1.7-3.3}$ &
        $70.0^{+0.7+1.3}_{-0.7-1.3}$ &
        $69.9^{+1.2+2.3}_{-1.2-2.3}$ &
        $69.9^{+0.5+1.3}_{-0.7-1.1}$\\
        
        $\Omega_{\rm m0}$ & 
        $0.287^{+0.020+0.044}_{-0.024-0.042}$ &
        $0.291^{+0.007+0.016}_{-0.008-0.014}$&
        $0.291^{+0.016+0.030}_{-0.016-0.029}$ &
        $0.291^{+0.007+0.014}_{-0.007-0.014}$\\
        
        $\Omega_{\sigma 0}$ (95$\%$ C.L.) &
        $<2.93 \times 10^{-15}$ &
        $<3.04 \times 10^{-15}$ &
        $<2.81 \times 10^{-15}$ &
        $<2.72 \times 10^{-15}$ \\

        $\ln \mathcal{E}$ & 
        $-20.45 \pm 0.13$ & $-24.32 \pm 0.14$ &
        $-539.99 \pm 0.18$  & 
        $-543.74 \pm 0.18$ 
        \\
        
        \hline
        \hline

         $\Lambda$CDM & & &\\
        \hline
        $H_0$ &
        $70.2^{+1.7+3.3}_{-1.7-3.2}$ &
        $69.4^{+0.6+1.1}_{-0.6-1.1}$&
        $69.5^{+1.1+2.3}_{-1.1-2.1}$&
        $69.3^{+0.5+1.0}_{-0.5-1.0}$
        \\

        $\Omega_{\rm m0}$ &
        $0.280^{+0.020+0.046}_{-0.024-0.040}$ &
        $0.290^{+0.008+0.015}_{-0.007-0.014}$ &
        $0.288^{+0.016+0.030}_{-0.016-0.030}$ &
        $0.291^{+0.007+0.014}_{-0.007-0.014}$ 
        \\
        
        $\ln \mathcal{E}$ & 
        $-18.1 \pm 0.12$ & $-22.1 \pm 0.17$ &
        $-537.6 \pm 0.30$ & 
        $-541.91 \pm 0.17$ \\
        
        \hline
        \hline
        
    \end{tabular}%
    }

\end{table*}

\begin{figure}[hbt!]
	\centering
	\includegraphics[width=8.5cm]{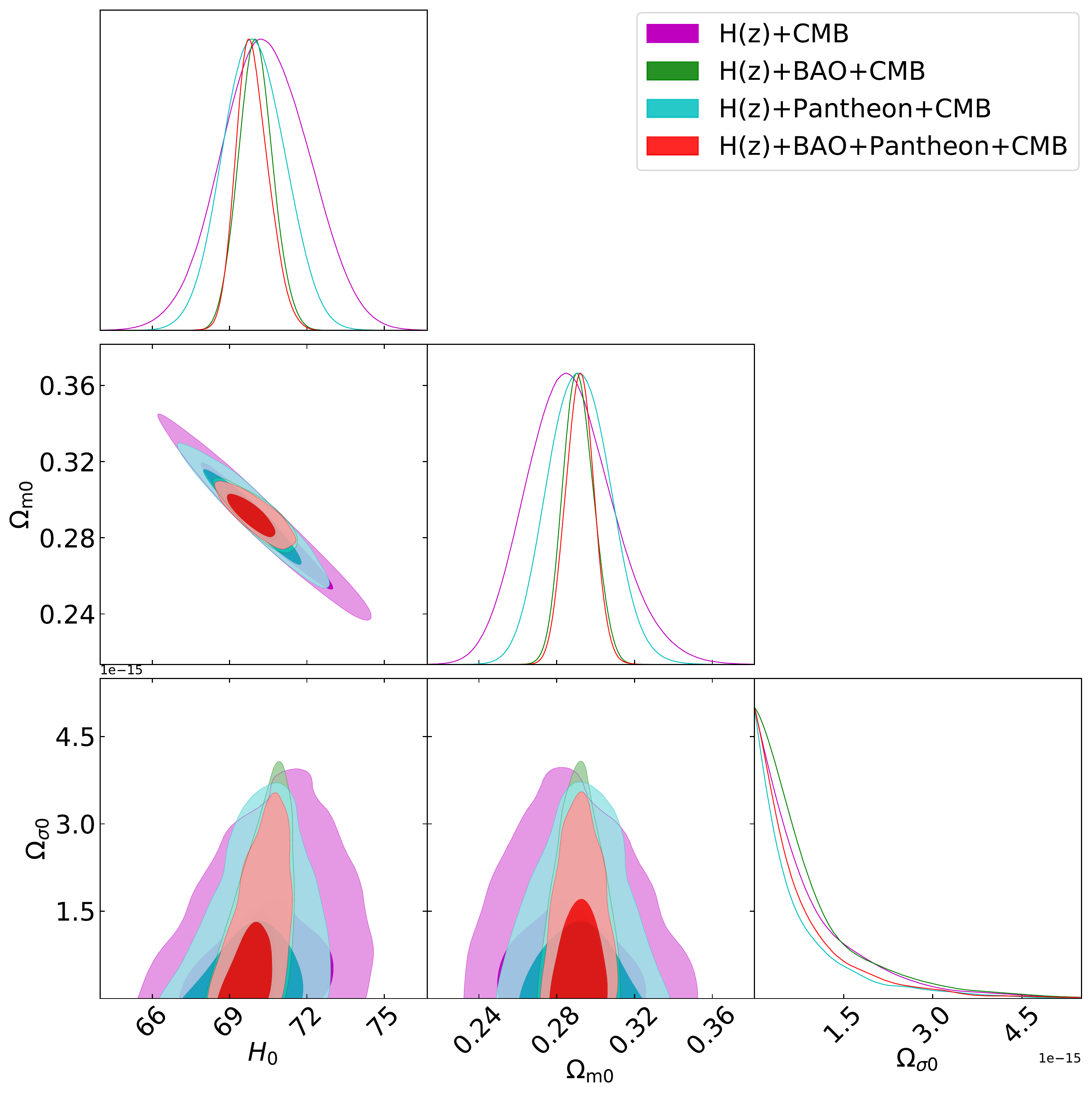}
	\caption{One-dimensional and two-dimensional marginalized confidence regions (68\% and 95\%  C.L.) for the anisotropic $\Lambda$CDM model parameters $H_0$, $\Omega_{\rm m0}$ and $\Omega_{\sigma 0}$ from $H(z)$, CMB, BAO and Pantheon data combinations.}
	\label{fig:comb}
\end{figure}

\begin{figure*}[htb!]
	\centering
	\includegraphics[width=8.5cm]{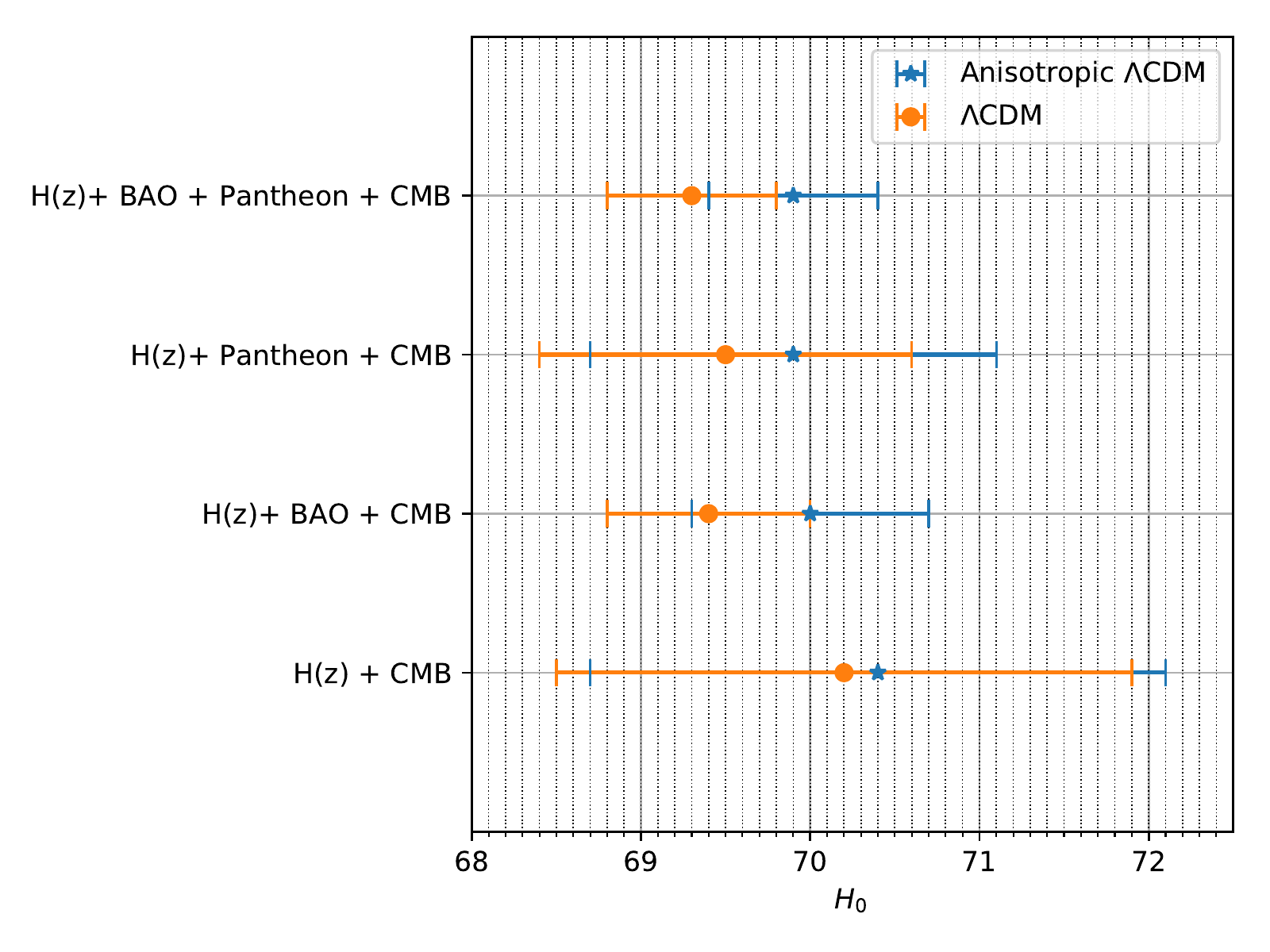}
	\includegraphics[width=8.5cm]{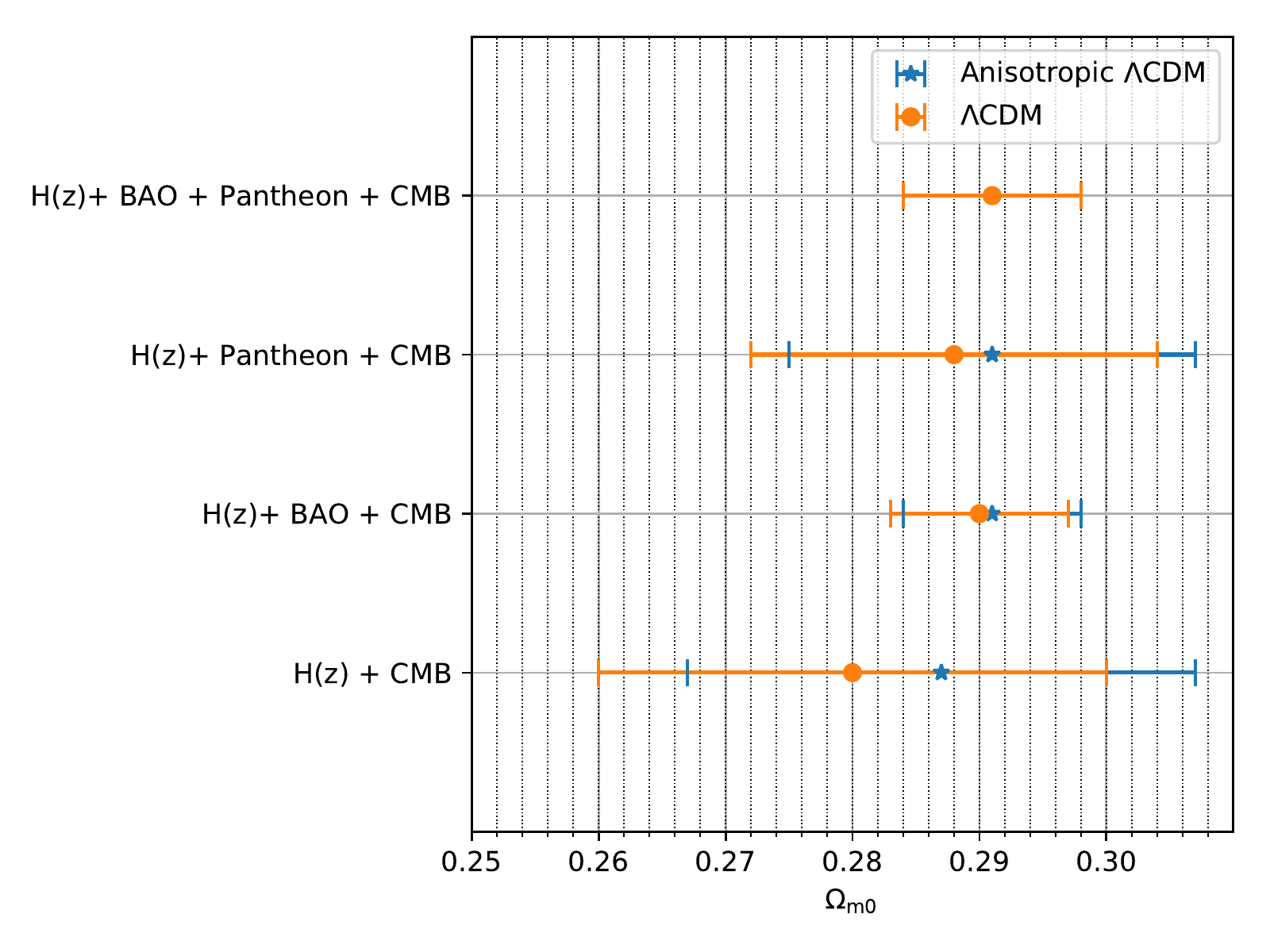}
	\caption{68\% confidence intervals of $H_0$ and $\Omega_{\rm  m0}$ for the anisotropic $\Lambda$CDM in comparison with the $\Lambda$CDM model.}
	\label{fig:H0om0}
\end{figure*}

 Figure \ref{fig:comb} shows the one-dimensional and two-dimensional (68\% and 95\%) confidence regions of the anisotropic $\Lambda$CDM model parameters for four different data combinations, each including the CMB data.
From Table \ref{comb_results} and Fig. \ref{fig:comb}, one may see that the joint dataset $H(z)$ + BAO + Pantheon + CMB yields the most tight constraints on all the model parameters. Further, in the anisotropic $\Lambda$CDM model we notice constraints on $H_0$ and $\Omega_{\rm m0}$ similar to the $\Lambda$CDM model. 
 We observe that the mean values of $H_0$ and $\Omega_{\rm m0}$ in the case of the anisotropic $\Lambda$CDM model are systematically larger than those in the case of the standard $\Lambda$CDM model, though not significantly. One may visualize the precise shift in these parameters due to anisotropy in Fig.\ref{fig:H0om0}, where the 68\% C.L. ranges of $H_0$ and $\Omega_{\rm  m0}$ are displayed for the anisotropic $\Lambda$CDM in contrast with the $\Lambda$CDM model. In the anisotropic $\Lambda$CDM model in comparison to the $\Lambda$CDM model, 
 the predicted mean values of $H_0$ and $\Omega_{\rm m0}$ for different combinations of the datasets are more similar while the errors remain almost the same.
 Namely, the largest deviations in $H_0$ and $\Omega_{\rm m0}$ between the different combinations of datasets are $\Delta H_0=0.50$ and $\Delta\Omega_{\rm m0}=0.0048$ in the anisotropic $\Lambda$CDM, while those are approximately 2 times larger, $\Delta H_0=0.94$ and $\Delta\Omega_{\rm m0}=0.0113$, in the standard $\Lambda$CDM.
 
\begin{figure}[htb!]
	\centering
	\includegraphics[width=8.5cm]{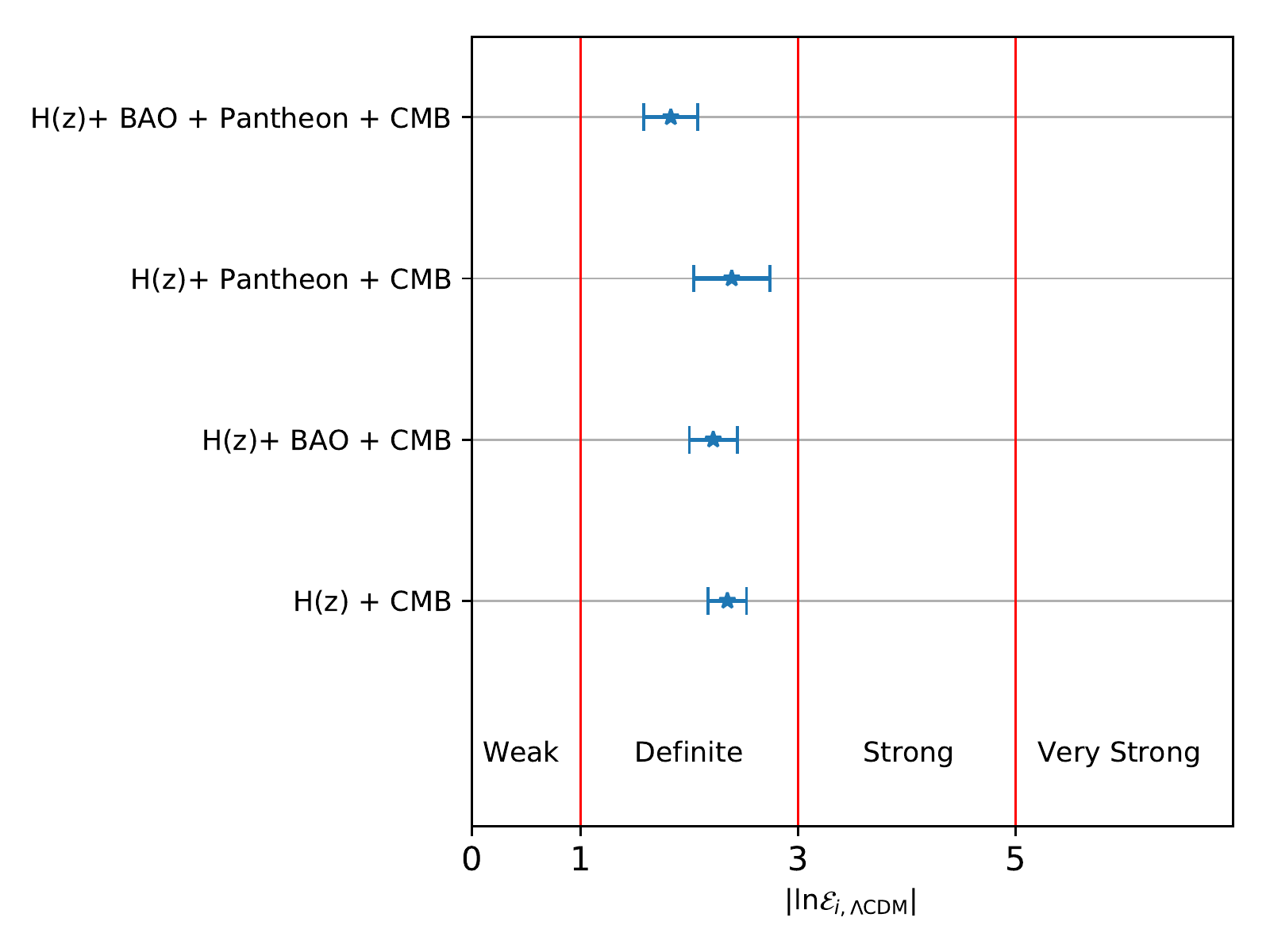}
	
	\caption{Summary of the Bayesian evidence for the anisotropic $\Lambda$CDM model in comparison with the $\Lambda$CDM model.}
	\label{fig:comb_be}
\end{figure}
 
Figure \ref{fig:comb_be} shows a summary of the Bayesian evidence of the anisotropic $\Lambda$CDM model in comparison with the $\Lambda$CDM model. We observe definite evidence in all cases of the data combinations as per the Jeffrey's scale in Table \ref{tab:BE}.

\section{Further Discussions}\label{S5}
In this section, we further discuss our results in the context of matter-radiation equality, big bang nucleosynthesis and CMB quadrupole problem.

\subsection{Matter-radiation equality}
\label{sec:MREq}
The transition from radiation domination to matter domination is one of the most important epochs in the history of the Universe. This transition alters the growth rate of density perturbations: during the radiation era perturbations well inside the horizon are nearly frozen but once matter domination commences, perturbations on all length scales are able to grow by gravitational instability and therefore it sets the maximum of the matter power spectrum \cite{Liddle:1998ij}. Namely, it determines the wave number, $k_{\rm eq,m,r}$, of a mode that enters the horizon, $H_{\rm eq,m,r}a_{\rm eq,m,r}$, at the matter-radiation transition \cite{Liddle:1998ij}. In our model, $k_{\rm eq,m,r}=H_{\rm eq,m,r}a_{\rm eq,m,r}$ can be estimated analytically by assuming $H(a)$ given in \eqref{model} that holds all the way to matter-radiation equality. At this point, both the radiation and matter contribute equally to the total energy density. At radiation-matter equality, the wave number of a mode that enters the horizon is given by
$k_{\rm eq,m,r}=a_{\rm eq,m,r}H_{\rm eq,m,r}.$
In the anisotropic $\Lambda$CDM model, it turns out that
\begin{equation}
\begin{aligned}
\frac{k^2_{\rm eq,m,r}}{H_0^2}
=&\,\Omega_{\sigma 0}(1+z_{\rm eq,m,r})^4 + 2\Omega_{\rm m0}(1+z_{\rm eq,m,r})\\&+\Omega_{\Lambda 0}(1+z_{\rm eq,m,r})^{-2},
\end{aligned}
\end{equation}
where $\Omega_{\Lambda0}=1-2\Omega_{\rm m0}-\Omega_{\sigma 0}$. The combined data [$H(z)$+BAO+Pantheon+CMB], from our observational analysis, predict the matter-radiation equality redshift as
\begin{equation}
z_{\rm eq,m,r}=-1+\frac{\Omega_{\rm m0}}{\Omega_{\rm r0}}=3412^{+45}_{-59},
\end{equation}
and
\begin{equation}
k_{\rm eq,m,r}=0.01040\pm 0.00010\,{\rm Mpc}^{-1},
\end{equation}
which are pretty much consistent with the recent Planck results; for instance, Planck TT+lowE gives $z_{\rm eq,m,r} =3411\pm48$ and $k_{\rm eq,m,r} = 0.01041\pm0.00014$ Mpc$^{-1}$  \cite{Aghanim:2018eyx}, implying that the matter perturbations in the anisotropic $\Lambda$CDM model are not affected by the expansion anisotropy.

\subsection{Big bang nucleosynthesis}

BBN \cite{Kneller:2004jz,Steigman:2007xt} provides a probe of the dynamics of the early Universe, which in turn would give us an opportunity to further investigate the constraints on the anisotropic $\Lambda$CDM model. Such that, in the standard-BBN model---assuming that the standard model of particle physics is valid, and the expansion of the Universe is governed by GR---the processes relevant to BBN take place during the time evolution of the Universe from $t\sim 1\,{\rm s}$ to $t\sim 3\,{\rm min}$ corresponding to the temperature change from $T\sim 1\, {\rm MeV}$ to $T\sim 0.1\, {\rm MeV}$, during the radiation-dominated era. This scenario, of course, should not be altered significantly in a viable cosmological model.
The radiation-expansion anisotropy equality redshift $z_{\rm eq,\sigma,r}$ is evaluated as
\begin{equation}
z_{\rm eq,\sigma,r}=-1+\left(\frac{\Omega_{\rm r0}}{\Omega_{\sigma0}}\right)^\frac{1}{2}.
\end{equation}
Also, $z_{\rm BBN} \sim 3 \times 10^{8}$ is the redshift of the physical processes relevant to BBN that takes place in standard cosmology. Therefore, it would then roughly require $z_{\rm eq,\sigma,r}>z_{\rm BBN}$ in order to avoid any possible adverse effects on the BBN phenomenon due to expansion anisotropy. This inequality leads to
\begin{equation}
 \Omega_{\sigma0} < \frac{\Omega_{\rm r0}}{(z_{\rm BBN}+1)^2}, 
\end{equation}
and given that $\Omega_{\rm r0}\sim10^{-4}$, the upper bound of $\Omega_{\sigma0}$ yields
$\Omega_{\sigma0}< 10^{-21}$, that lies within the probability region of $\Omega_{\sigma0}$ given by the constraint $\Omega_{\sigma0}< 10^{-15}$ (see Table \ref{comb_results}). 


On the other hand, as may be seen from the investigations in Refs. \cite{Barrow:1976rda,Campanelli:2011aa}, the expansion anisotropy would not lead to a considerable deviation from the standard BBN for the  $\Omega_{\sigma}/\Omega_{\rm r}$ ratio up to a few percent, viz., \begin{equation}
    \frac{\Omega_{\sigma}(z=z_{\rm BBN})}{\Omega_{\rm r}(z=z_{\rm BBN})}\lesssim 10^{-2}.
\end{equation}
Considering this, the upper bound on $\Omega_{\sigma0}$ further improves to $\Omega_{\sigma0}\lesssim 10^{-23}$. Thus, we see that BBN offers tight constraints on the anisotropy parameter $\Omega_{\sigma0}$ in comparison to the constraint $\Omega_{\sigma0}< 10^{-15}$ obtained here directly using the cosmological data. However, it may be noted that the constraint on $\Omega_{\sigma0}$ from BBN may be weaker than the one obtained from the cosmological data in the presence of anisotropic sources.

\subsection{CMB quadrupole problem}

The CMB power spectrum at $l=2$ (quadrupole) corresponds to the angular scale $\theta=\pi/2$ on the sky ($\ell=\frac{\pi}{\theta}$). Corresponding to $l=2$, the observed value of the temperature fluctuation by Planck is  $\Delta T_{\rm Planck} \approx 14 \,\mu K$ \cite{Ade:2013kta}, whereas the standard $\Lambda$CDM predicted value reads $\Delta T_{\rm st}\approx 34\, \mu K$. Clearly, there is a considerable discrepancy between the two values. This deficit can be reduced partially by using cosmic variance, viz., $\Delta T_{\rm st+variance} \approx 28 \,\mu K$ \cite{Dodelson03,Campanelli:2006vb}.
Here, we look for the possible effects of the expansion anisotropy on the CMB power spectrum. Anisotropic expansion of the Universe implies a different evolution of the temperature of the free streaming photons for the different expansion factors in three orthogonal directions. 
This in turn would alter the CMB power spectrum at $\ell=2$, without affecting the temperature fluctuations corresponding to higher multipoles as predicted within the standard $\Lambda$CDM model. 

We first set $x_1=0$ [viz., consider locally rotationally symmetric (LRS) anisotropy \cite{Ellis:1998ct} for convenience] and thereby find $x_2=-x_3$ leading to $\sigma_0^2=x_3^2/3$ from \eqref{2.17}. For small anisotropies, the evolution of the photon temperature along the $x$ and $z$ axes can be given as \cite{Barrow1985,Barrow:1997sy}
\begin{align}
T_{x} &=&T_0\frac{a_{x0}}{a_x}=T_0 e^{-\int H_x {\rm d} t}\simeq {T_0}-{T_0}\int H_x {\rm d} t,  \label{eq.L1}\\
T_{z} &=&T_0\frac{a_{z0}}{a_z}=T_0 e^{-\int H_z {\rm d} t}\simeq {T_0}-{T_0}\int H_z {\rm d} t,  \label{eq.L2}
\end{align}
where today's CMB monopole temperature is $T_0 =2.7255 \pm 0.0006 \,{\rm K}$ \cite{Fixsen09}. In the case of LRS anisotropy, using \eqref{2.17} and \eqref{2.7a}, the temperature difference along the $x$ axis and $z$ axis can be given as 
\begin{equation}
\begin{aligned}
T_x-T_z&=T_0\int_{t_{\rm rec}}^{t_0}(H_x-H_z){\rm d}t =T_0 \int_{t_{\rm rec}}^{t_0}\sqrt{3} \sigma {\rm d}t\\
&=3T_0\sqrt{\Omega_{\sigma0}}\int_0^{z_{\rm rec}}\frac{H_0(1+z)^2}{H(z)}{\rm d}z.
\label{deltaT}
\end{aligned}
\end{equation}
Using the upper bound $\Omega_{\sigma0}\sim 10^{-15}$ and mean values of other parameters, we obtain $\Delta T_{\Omega_{\sigma}}= T_x-T_z\sim 10.72\;\rm mK$, whereas, we need $\Omega_{\sigma0}\sim 4\times 10^{-20}$ to get $\Delta T_{\Omega_{\sigma}}\sim 20 \,\mu K$, which could be considered for addressing the quadrupole temperature problem. For, in the latter case, provided that the orientation of the expansion anisotropy is set properly, it is possible to reduce $\Delta T_{\rm st}\approx 34 \,\mu K$ in $\Lambda$CDM model to the observed value $\Delta T_{\rm Planck} \approx 14 \, \mu K$ \cite{Ade:2013kta}. However, the strict upper bound $\Omega_{\sigma0}\sim 10^{-23}$ from BBN, yields $\Delta T_{\Omega_{\sigma}}\sim 1\;\mu \rm K$. Thus, the BBN constraint on $\Omega_{\sigma0}$ prohibits significant modification in the quadrupole temperature.




\section{Conclusions}
\label{sec:Conclusions}

We have considered the simplest anisotropic generalization, as a correction, to the standard $\Lambda$CDM model, constructed by replacing the spatially flat RW metric by the Bianchi type-I metric while keeping all other constituents (e.g., the physical ingredients of the Universe) of the model as usual. This modifies the Friedmann equation of the standard $\Lambda$CDM model by rendering $H(a)$ as the average expansion rate of the Universe, and including a new term $\Omega_{\sigma 0}a^{-6}$ (which mimics stiff fluid), where $a$ is the average expansion scale factor. We have carried out observational analysis by defining the corresponding average redshift $z$ through the mean scale factor $a$ as $z=-1+\frac{1}{a}$. Accordingly, we should note that the constraints obtained here on $\Omega_{\sigma 0}$, viz., the present day expansion anisotropy, are model dependent. The method employed here constrains the stiff-fluid-like effect of the expansion anisotropy, which arises within GR in the absence of any kind of anisotropic source, in the model under consideration. From the data [$H(z)$ and Pantheon] relevant to the late Universe ($z\lesssim 2.4$) only, we have obtained the constraint $\Omega_{\sigma0}\lesssim10^{-3}$. This being comparable with the model-independent constraints \cite{Campanelli:2010zx,Wang:2017ezt} shows that the method we employed is useful and informative. When the BAO and CMB data also are included in our analysis, the constraint improves by 12 orders of magnitude, i.e., $\Omega_{\sigma0}\lesssim10^{-15}$. The reason for such a tight constraint is that we guaranteed expansion anisotropy to remain as a correction all the way to the largest redshifts involved in BAO and CMB data used in our analyses, by fixing the drag redshift as $z_{\rm d}=1059.6$ (involved in BAO analysis) and the last scattering redshift as $z_* = 1089.9$ (involved in CMB data), where the fixed values are taken from  the Planck 2015 release for the standard $\Lambda$CDM model \cite{Ade}. In all cases of the dataset combinations, the Bayesian evidence of the anisotropic $\Lambda$CDM model in comparison with the $\Lambda$CDM model is definite as per the Jeffrey's scale. We have found that, in comparison with the standard $\Lambda$CDM model, $\Omega_{\sigma 0}\lesssim10^{-15}$ could alter neither the matter-radiation redshift $z_{\rm eq,m,r}$ nor the peak of the matter perturbations $k_{\rm eq,m,r}$. Finally, demanding that the expansion anisotropy has no significant effect on the standard BBN, we have obtained a more tight constraint as $\Omega_{\sigma0}\lesssim10^{-23}$. 
We have shown explicitly that this constraint from BBN renders the expansion anisotropy irrelevant to make a significant contribution to the resolution of the quadrupole moment problem, whereas the constraint from the cosmological data in our model provides the temperature change up to $\sim11\, \rm mK$, which is much larger than the required $\sim20\, \mu \rm K$ temperature difference in the free streaming photons arriving us from the last scattering surface in the orthogonal directions in the sky. Such strong constraints, very much beyond the model-independent constraints, result from the data relevant to earlier Universe due to the quite steep redshift dependence of the shear scalar, $\sigma^2\propto (1+z)^6$. On the other hand, it is in principle possible to alter the redshift dependence of the shear scalar, particularly in the late Universe by replacing $\Lambda$ by an anisotropic DE (see, e.g., \cite{Tedesco} and references therein) or replacing GR by a modified gravity, such as Brans-Dicke theory, leading to effective anisotropic sources (see, e.g., \cite{Akarsu:2019pvi} and references therein). For example, it is possible to have cosmological models in which the shear scalar yields $\sigma^2\sim (1+z)^4$ as like radiation; then, in principle, it would contribute to the Friedmann equation like an extra relativistic degrees of freedom (e.g., neutrino). Accordingly, in such a setup, expansion anisotropy would play the role of the extra relativistic degrees of freedom in the $\Lambda$CDM model, and, given that the recent Planck release \cite{Aghanim:2018eyx} gives $\Delta N_{\rm eff}<0.30$, we would expect to obtain $\Omega_{\sigma0}\sim \Omega_{\rm r0}/10\lesssim10^{-6}$ from observational analysis, which is pretty much consistent with the direct observational constraints on the present day expansion anisotropy and it could also contribute to the quadrupole moment problem. Moreover, for the cosmological models in which the shear scalar yields a redshift dependence flatter than that of the radiation, we would expect the cosmological data to constrain $\Omega_{\sigma0}$ stronger than BBN. Thus, we conclude that constraining the present day expansion anisotropy (viz. $\Omega_{\sigma0}$) through the new term that appears in the Friedmann equation due to expansion anisotropy, as done in this paper, provides a very informative method and could successfully be used for constraining different kinds of anisotropic cosmological models and investigating their underlying theories in the light of data from cosmological observations.

\begin{acknowledgements}
 \"{O}.A. acknowledges the support by the Turkish Academy of Sciences in scheme of the Outstanding Young Scientist Award  (T\"{U}BA-GEB\.{I}P). S.K. and S.S. gratefully acknowledge the support from SERB-DST Project No. EMR/2016/000258, and DST Fund for Improvement of S$\&$T Infrastructure project No. SR/FST/MSI-090/2013(C). L.T. acknowledges partial support from the INFN-TAsP project and by the research Grant No. 2017W4HA7S ``NAT-NET: Neutrino and Astroparticle Theory Network'' under the program PRIN 2017 funded by the Italian Ministero dell'Istruzione, dell'Universit\`a e della Ricerca (MIUR).
\end{acknowledgements}

\end{document}